\documentstyle[12pt]{article}
\begin{document}
\renewcommand{\thepage}{ }
\begin{titlepage}
\title{
\hfill
\vspace{1.5cm}
{\center Magnetic field relaxation in ferromagnetic Ising systems}
}
\author{
R. M\'elin\\
{}\\
{CRTBT-CNRS, 38042 Grenoble BP 166X c\'edex France}}
\date{}
\maketitle
\begin{abstract}
\normalsize
We analyze the thermal magnetization reversal processes in magnetic grains.
Two experiments are carried out: swtiching time and switching field
experiments. In both cases, we find that the simulated behavior is
coherent with existing experimental data (the streched exponent of the
switching time experiment increases with the temperature and is
superior to unity; there exists a master curve for the switching field
experiment). Moreover, we simulated magnetic grains in a region of
parameters where no experimental data are available. We find that
the relaxation time distribution $P(\ln{\tau})$ is gaussian, and we find the
existence of a strong field regime.
\end{abstract}
\end{titlepage}

\newpage
\renewcommand{\thepage}{\arabic{page}}
\setcounter{page}{1}
\baselineskip=17pt plus 0.2pt minus 0.1pt
\tableofcontents

Magnetic grains are considered as good candidates to bring to the fore
macroscopic quantum tunelling \cite{S C B}. Before considering
the problem of quantum tuneling at low temperatures, it would first be
interesting to understand thermally activated processess at higher temperature.
The aim of the present paper is to address this question in a simplified model.
Experimentally, small ferromagnetic grains were studied for many years in
powders, so that data were available only for a collection of
interacting magnetic grains, with different sizes and compositions.
Using magnetic force microscopy \cite {MFM} and Lorentz microscopy
\cite{Lorentz}, it was possible to resolve the magnetic properties
of individual single-domain particles. The present article is based
on the experimental works \cite{exp}, where it was possible to study
the magnetization reversal of a single monodomain magnetic particule
of size
$0.1 \mu m$ x $0.05 \mu m$. Due to dipolar interactions,
anisotropy aligns the spins along the largest direction of the grain.
The spins are essentially XY spins since the grain is essentially
bidimensional.
We do not carry out direct simulations of XY spins at finite temperature,
but, instead, we consider a simplified Ising spin model on a disk to
simulate a magnetic grain. This means that a direct comparison with
the experimental exchange couplings and magnetic field is not possible.
However, we believe that our simulations are qualitatively valid, eventhough
they do not directly represent magnetic grains.
The reason why we choose the disk geometry is that the frontier is expected
to play a great role in the magnetization reversal processess.
In \cite{Richard}, the boundary conditions are periodic, so that boundary
effects are not properly taken into account.
In order to compare with existing experiments \cite{exp}, we use the same
experimental techniques than in \cite{exp}.
Two types of experiments are possible 1) switching time experiments
2) switching field experiments.

In the first class of simulations, the
system starts in a low temperature configuration where nearly all the
spins are paralell. A constant magnetic field opposite to the initial
magnetization is applied from time $t=0$. The initial configuration,
antiparalell to the field, is then metastable, and the magnetization reverses
to the oppposite magnetization. The time $\tau$ corresponding to the
cancellation of the magnetization is recorded. Since the reversal of
magnetization process is stochastic, $\tau$ is a random variable.
In our simulations, we took into account 30000 different runs. This
number is necessarily huge in order to have a good statistics.
In existing experiments, data with such an important number of
reversal processes are not available at the moment, so that some
of the results that are reported here are not directly comparable
to experimental data because of the lack of experimental data in these
temperature and field ranges.
We first show that the distribution $P(\ln{\tau})$ of the logarithm
of the relaxation times is gaussian. This result is in agreement with
other theories, especially for spin glasses \cite{Souletie}, where a
gaussian distribution of the logarithm relaxation times has been assumed,
in the context of a phenomenological renormalization.
We also analyze the mean value $\langle \ln{\tau}
\rangle$, and the variance $\langle ( \ln{\tau} - \langle \ln{\tau}
\rangle)^{2} \rangle$ as a funtion of temperature. We find that, if the
field is not too strong, both the mean value and the variance diverge
as $T$ approaches zero. If the field is too strong, the divergence is replaced
by a saturation of these two quantities, so that long relaxation
times at low temperature are inhibited by a too strong magnetic field.
This qualitative behavior between a strong magnetic field regime and a
weaker magnetic field regime is to be compared with the results of
\cite{Richard}, where a strong magnetic field regime is exhibited.
In the switching time experiments, we can study the variations of
$\langle \ln{\tau} \rangle$ as a function of the variance
$\langle ( \ln{\tau} - \langle \ln{\tau} \rangle)^{2} \rangle$.
Within the context of phenomenological spin glass theories \cite{Souletie},
this curve is expected to be linear. However, we do not find linear
variations. Depending on the strength of the magnetic field, the
curvature is negative (strong magnetic fields) or negative (weak magnetic
fields).

The second possible class of experiments is the so-called switching
filed experiments.
The idea is to start from a low temperature configuration,
and to switch a field oppposite to the global magnetization.
The field increases linearly with a rate $\nu = -d H/dt$.
In the presence of the external field, the magnetization reverses
and one measures the magnetic field $H_{sw}$ for which the
magnetization cancels. The switching field $H_{sw}$ is averaged over
a large number of runs, and computed for various temperatures.
In order to analyze the data, we use the phenomenological model
of reference \cite{Barbara}. We find that, as in real experiments,
there exists a master curve for  $\langle H_{sw} \rangle(T,\nu)$.

\section{Switching time experiment}
\subsection{Square lattice with free boundary conditions}
In order to emphasize the importance of the boundary conditions,
we first consider the case of an Ising model on a square.
In this case, the domains of reversed spins can nucleate from the
corner. The barrier to nucleate a domain of spins paralell to the
magnetic field is $E_1(R) = \pi R(J - h R/2)$, where we have assumed
that the domain of reversed spins is circular (this is an approximation, since
the droplets are not necesserily circular). The maximum over $R$ of
$E_1(R)$ is nothing but the barrier that need to be paid to
reverse the magnetization. This barrier is $E_1(R_1^{*}) = \pi J^{2}/2 h$.
In order to compare with the absence of edges (Ising model on a disk),
we calculate the barrier to nucleate a domain of spins paralell
to the field on the disk. We assume that the radius of the disk
is large compared to the radius of curvature of the droplet.
The barrier is given by $E_2(R) = 2 \pi R(2J-hR)$.
The first term $4 \pi R J$
is the energy term associated to the antiparallel links at the boundary
and $-2 \pi h R^{2}$ is the magnetic energy. $E_2(R)$ is maximum for
$R_2^{*} = J/h$, which corresponds to the critical size of the domain.
If $R<R_2^{*}$, the domain has a trend to regress whereas it naturally
grows if $R>R_2^{*}$. The barrier height is $E_2(R_2^{*}) = 2 \pi J^{2}/h$.
We see that the barrier in the presence of edges is one half of the
barrier in the absence of edges, from what
we conclude that the spin system on a square with open boundary
conditions is very sensitive
to the presence of edges and will first flip from the edges, at least
if the magnetic field is not too strong.
To check this assertion, we took a snapshot of the system during a given run.
The result is pictured on figure \ref{fig1}.
It is clear that the edges play an important role in the magnetization
reversal, since
the droplets are pinned at the edges, where the energy barrier is
smaller than in the bulk. The pinning of the droplets is here purely
geometric.
One could also imagine that the droplets are pinned due to the
presence of impurities.

In what follows, we aim to elimitate the corner nucleation of domains.
This is why we consider an Ising model on a disk, with free boundary
conditions.

\subsection{Distribution $P(\ln{\tau})$}
We call $\tau$ the time corresponding to the cancelation of the magnetization.
We plotted on figure \ref{fig2} the distribution $P(\ln{\tau})$ as
a function of $\ln{\tau}$. We see that these distribution in very good
agreement with gaussian distributions. This means that the distribution
of relaxation times in an external magnetic field is gaussian, provided
one uses the logarithm of the relaxation times.
This result is in agreement with the assumption of \cite{Souletie}.
In the context of spin glasses, it is assumed that $P(\ln{\tau})$ is
Gaussian. However, as we will see later, not all the assumptions
of this phenomenological theory are valid in our model.

\subsection{Mean value and variance of the relaxation time distribution}
Since $P(\ln{\tau})$ is a well defined gaussian distribution, we can
study the variations of $\langle \ln{\tau} \rangle$ and of the variance
$\langle ( \ln{\tau} - \langle \ln{\tau} \rangle)^{2} \rangle$
as a function of
temperature. We first begin with the mean value $\langle \ln{\tau}
\rangle$. The results for three different magnetic fields are plotted
on figure \ref{fig3}. We observe that if the intensity of the
magnetic field decreases, the relaxation times increase. Moreover, if there
exists a critical value $h^{*}$ f the magnetic field such that if $|h|<h^{*}$,
the relaxation times diverges in the zero temperature limit, whereas
if $|h|<h^{*}$, the relaxation times saturate in the zero temperature
limit. We identify this behavior with the existence of the strong
magnetic field regime of reference \cite{Richard}.

The variations of the variance $\langle (\ln{\tau} - \langle \ln{\tau}
\rangle )^{2} \rangle$ as a function of temperature are plotted
on figure \ref{fig4}.
These variations
are qualitatively similar to the variations of the mean value:
if the intensity of the magnetic field is too strong, we observe a
saturation of the variance, whereas it diverges for weaker
magnetic fields.

We are now in position to study the variations of the mean value
$\langle \ln{\tau} \rangle$
as a function of the variance
$\langle ( \ln{\tau} - \langle \ln{\tau} \rangle)^{2} \rangle$.
These variations are pictured on figure \ref{fig5}.
We observe that in the strong magnetic field region, the variance
and the mean value are quite small, and both tend to a constant
if the temperature decreases to zero. By contrast, we observe that
if $h=-1.5$, the curvature is negative, whereas it is positive
if $h=-1$. It would be interesting to know whether experimental on
magnetic grains could confirm these simulations. However, these
data are not available at the moment, due essentially to the fact that
it is necessery to repeat the experience a large number of times to
obtain a good statistics, and also to the fact that the experiments
are not carried out for a given magnetic field, in order to avoid very
long relaxation times.
If we compare our results to the prediction of spin glass theory
\cite{Souletie},
we do not find that $\langle \ln{\tau} \rangle$ is linear
as a function of $\langle ( \ln{\tau} - \langle \ln{\tau}
\rangle)^{2} \rangle$, and, according to the strength of the magnetic field,
we find a positive or a negative curvature.
We conclude that our model for thermal magnetic grains shares
some common features with spin glasses
since $P(\ln{\tau})$ is found to be
gaussian in both cases.
However, whereas spin glasses phenomenological
models predict a linear
variation of $\langle \ln{\tau} \rangle$ as a function of
the variance $\langle ( \ln{\tau} - \langle \ln{\tau}
\rangle)^{2} \rangle$, we do not find here such a linear behavior.

\subsection{Streched exponent $\beta$}
An other possible analysis of the data of the switching time experiment
if to buid the histogram $P(\tau)$, and to fit the integrated histogram by
a streched exponential
\begin{equation}
\label{eq2}
\int_0^{\tau} P(\tau') d \tau' \simeq
1 - \exp{\left(-\left(\frac{\tau}{\tau_0} \right)^{\beta}\right)}
{}.
\end{equation}
This analysis is carried out on experimental data in \cite{exp}, where
the streched exponent $\beta$ is studied as a function of temperature.
The main results are that $\beta$ increases with temperature, and may be
superior to unity. As we shall see, our simulations agree with these two
experimental facts. At this point, it should be underlined that the
streched exponent $\beta$ has nothing to do with the streched exponent of
\cite{Ogielski} in the context of spin glasses. In \cite{Ogielski},
the autocorrelation functions of spin glasses are fitted by a streched
exponential, and a streched exponent is deduced from the numerical data.
In this case, the streched exponent is necessery inferior to unity,
whereas in what we study, the streched exponent may be larger than unity.

We show on figure \ref{fig6} the variation of the streched exponent
$\beta$ as a function of $T$, for two magnetic fields. It is clear
that $\beta$ increases with $T$, and may be superior to unity.
We also observe that the streched exponent increases if the strength
of the magnetic field increases, which is coherent with the fact
both the mean value $\langle \ln{\tau} \rangle$
and the variance $\langle ( \ln{\tau} - \langle \ln{\tau} \rangle)^{2}
\rangle$ decrease if the strengh of the magnetic field increases.

\section{Switching field experiment}
In the switching field experiment, a linearly increasing magnetic field
is applied opposite to the direction of the initial magnetization,
with a rate $\nu=-d H/dt$. The magnetic field $H_{sw}$ corresponding to the
cancellation of the magnetization is measured. In order to analyze
the data, we follow reference \cite{exp}, and use the phenomenological model
of \cite{Barbara}. In this model, the average switching field
$\langle H_{sw} \rangle$ is given by
\begin{equation}
\label{eq4}
\langle H_{sw} \rangle = H^{0}_{sw} \left( 1 - f_{\alpha}(T,\nu) \right)
,
\end{equation}
with
\begin{equation}
f_{\alpha}(T,\nu)=
\left(
\frac{k_B T}{E_0} \ln{\left(\frac{c T}{\nu \epsilon^{\alpha-1}}\right)}
\right)^{1/\alpha}
\end{equation}
$H_{sw}^{0}$, $E_0$ and $c$ are some constants, $\alpha=2$ for an
ideal single domain model and
\begin{equation}
\epsilon=(1-H_{sw}/H_{sw}^{0})^{1/\alpha}
{}.
\end{equation}
The experimental data consist of $\langle H_{sw} \rangle(T,\nu)$
as a function of the temperature $T$ and the switching rate $\nu$.
The relation (\ref{eq4}) predicts the existence of
a master curve for $\langle H_{sw} \rangle(T,\nu)$.
We carried out numerical simulations for the square lattice on the
disk, with open boundary conditions. We come to the conclusion
that there exists a master curve for $\langle H_{sw} \rangle(T,\nu)$.
The master curve is plotted on figure \ref{fig7}. We see that there
exists differences between the master curve and the behavior
$\langle H_{sw} \rangle/H_{sw}^{0} = 1-f_{\alpha}(T,\nu)$.
These deviations also exist in the experiments \cite{exp}.

\section{Conclusion}
We have carried out numerical simulations of a simplified model of
magnetic grain in the thermal regime. We have carried two types
of numerical simulations: switching time and switching field experiments.
As far as switching time experiments are concerned, we found a gaussian
behavior for the histogram $P(\ln{\tau})$ of the logarithm of the
relaxation times. This behavior is compatible with existing theories
of slow relaxation phenomena. We find the existence of a strong
field regime, where both $\langle \ln{\tau} \rangle$
and the variance $\langle (\ln{\tau} - \langle \ln{\tau} \rangle)^{2} \rangle$
tend to a constant as the temperature approaches zero. This regime was
already predicted in \cite{Richard}, and called the strong field regime.
If the magnetic field is smaller, the mean value and the variance diverge
as the temperature approaches zero.
It would be interesting to study the magnetic field $h^{*}$,
which separates the strong magnetic field region, as a function of
the system size. It would also be interesting to carry out similar
simulation for non Euclidian boundaries, in order to have a more precise
idea of the importance of boundary effects. To do so, it is possible to
study the case of a hyperbolic lattice, or a Cayley tree, with a finite
fraction of the sites at the boundary even in the thermodynamic limit.
Unfortunately, experimental data on magnetic grains are not available
since these experiments require a large amount of data.
We also calculated the streched exponent $\beta$ as a function of $T$.
Like in the case of experiments, we find that $\beta$ may be superior to
unity, and that $\beta$ increases with $T$.

We also considered the switching field experiment. There exists a master
curve for the average of the switching field as a function of the switching
rate and the temperature, which is consistent with the experimental data.

The author acknowledges discussion with J.C. Angl\`es d'Auriac,
B. Barbara, B. Dou\c{c}ot, K. Hasselbach, J. Souletie
and W. Wernsdorffer.

\newpage
\renewcommand\textfraction{0}
\renewcommand
\floatpagefraction{0}
\noindent {\bf Figure captions}

\begin{figure}[h]
\caption{}
\label{fig1}
Snapshot of the spin configuration in a given numerical experiment,
in the case of a 90x90 square lattice. The magnetization is
equal to $3/4$.
The inverse temperature is $\beta=1.6 J$, and the applied magnetic field is
$h=- J$.
A symbol denotes a down spin (parallel to the field)
and the spin is up in the absence of symbol.
The droplets are pinned at the corners.
\end{figure}

\begin{figure}[h]
\caption{}
\label{fig2}
Distribution of $P(\ln{\tau})$ as a function of $\ln{\tau}$ for different
temperatures. The lattice is a disk with 2821 sites. The magnetic field
is set to $h=-1.5$ from time $t=0$. The different distribution and
the fit to the Gaussian are plotted for $T=0.22 J, 0.20 J, 0.18 J, 0.16J$.
30000 runs were carried out in each case.
\end{figure}

\begin{figure}[h]
\caption{}
\label{fig3}
Mean value $\langle \ln{\tau} \rangle$ as a function of temperature, for
three values of the magnetic field $h=-2 J$,$h=-1.5 J$ and $h=-J$.
The lattice is circular, with 2821 sites.
30000 runs were carried out in each case.
\end{figure}

\begin{figure}[h]
\caption{}
\label{fig4}
Variance $\langle (\ln{\tau} - \langle \ln{\tau} \rangle)^{2}
\rangle$ as a function of temperature, for
three values of the magnetic field $h=-2 J$,$h=-1.5 J$ and $h=-J$.
The lattice is circular, with 2821 sites.
30000 runs were carried out in each case.
\end{figure}

\begin{figure}[h]
\caption{}
\label{fig5}
Variance $\langle (\ln{\tau} - \langle \ln{\tau} \rangle)^{2}
\rangle$ as a function of the mean value $\langle \ln{\tau} \rangle$,
for three values of the magnetic field $h=-2 J$,$h=-1.5 J$ and $h=-J$.
The lattice is circular, with 2821 sites.
30000 runs were carried out in each case.
\end{figure}

\begin{figure}[h]
\caption{}
\label{fig6}
Variations of the streched exponent $\beta$ with the temperature,
and for $h=-1$ and $h=-1.5$.
The lattice is circular, with 2821 sites.
30000 runs were carried out in each case.
\end{figure}

\begin{figure}[h]
\caption{}
\label{fig7}
Master curve of the switching field experiment for the square
lattice on the disk.
We plotted $\langle H_{sw} \rangle/H_{sw}^{0}$ as a function
of $f_{\alpha}(T,\nu)$. The Boltzmann constant is equal to unity,
$\alpha=1.8$, $H_{sw}^{0}=2.4$,$E_0=5$ and $c=1/25$. On this curve, 7 series
of measures at 7 different temperatures are superposed.
The dashed line is the linear behaviour and corresponds to
$\langle H_{sw} \rangle/H_{sw}^{0} = 1-f_{\alpha}(T,\nu)$.
\end{figure}

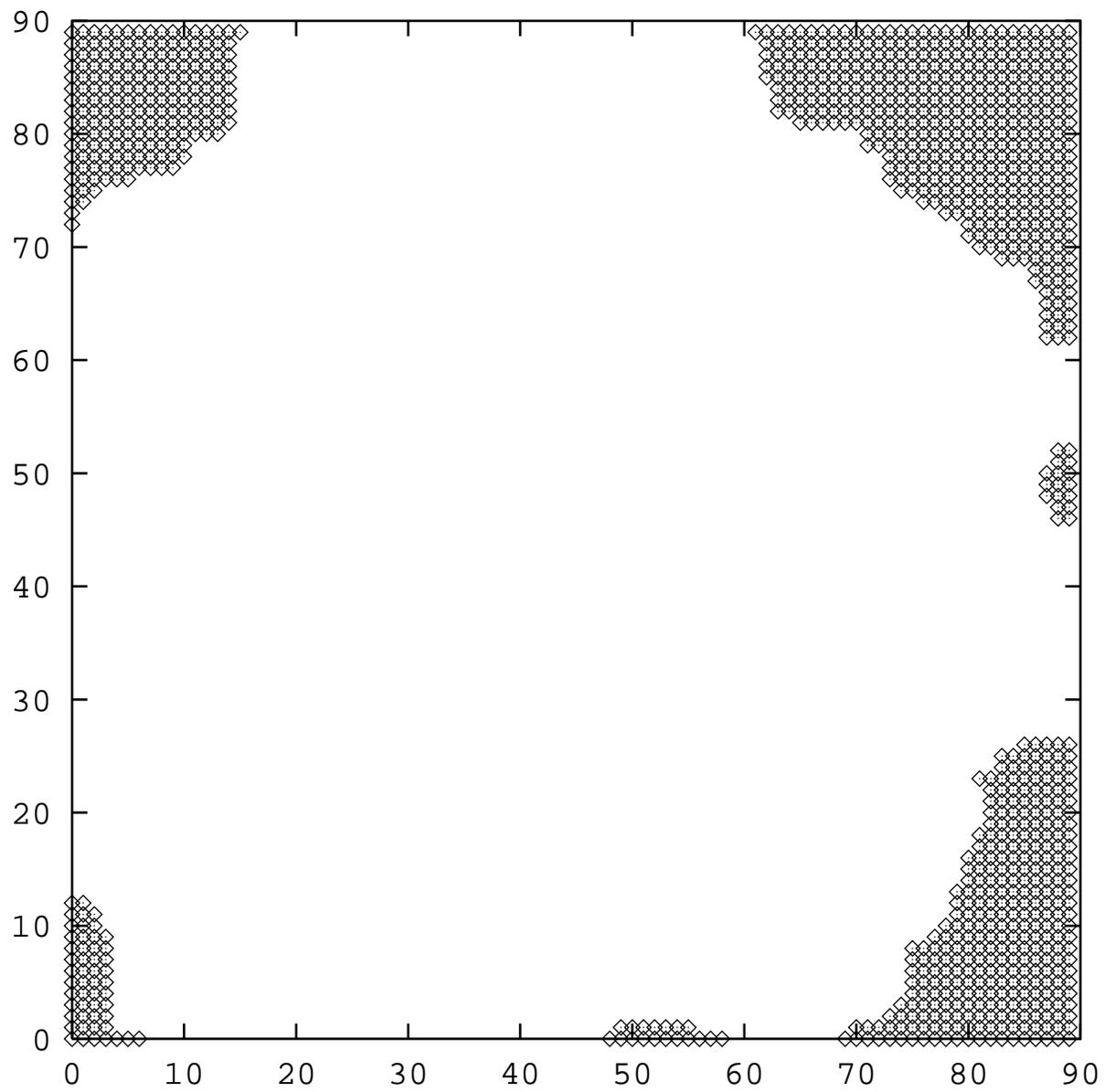
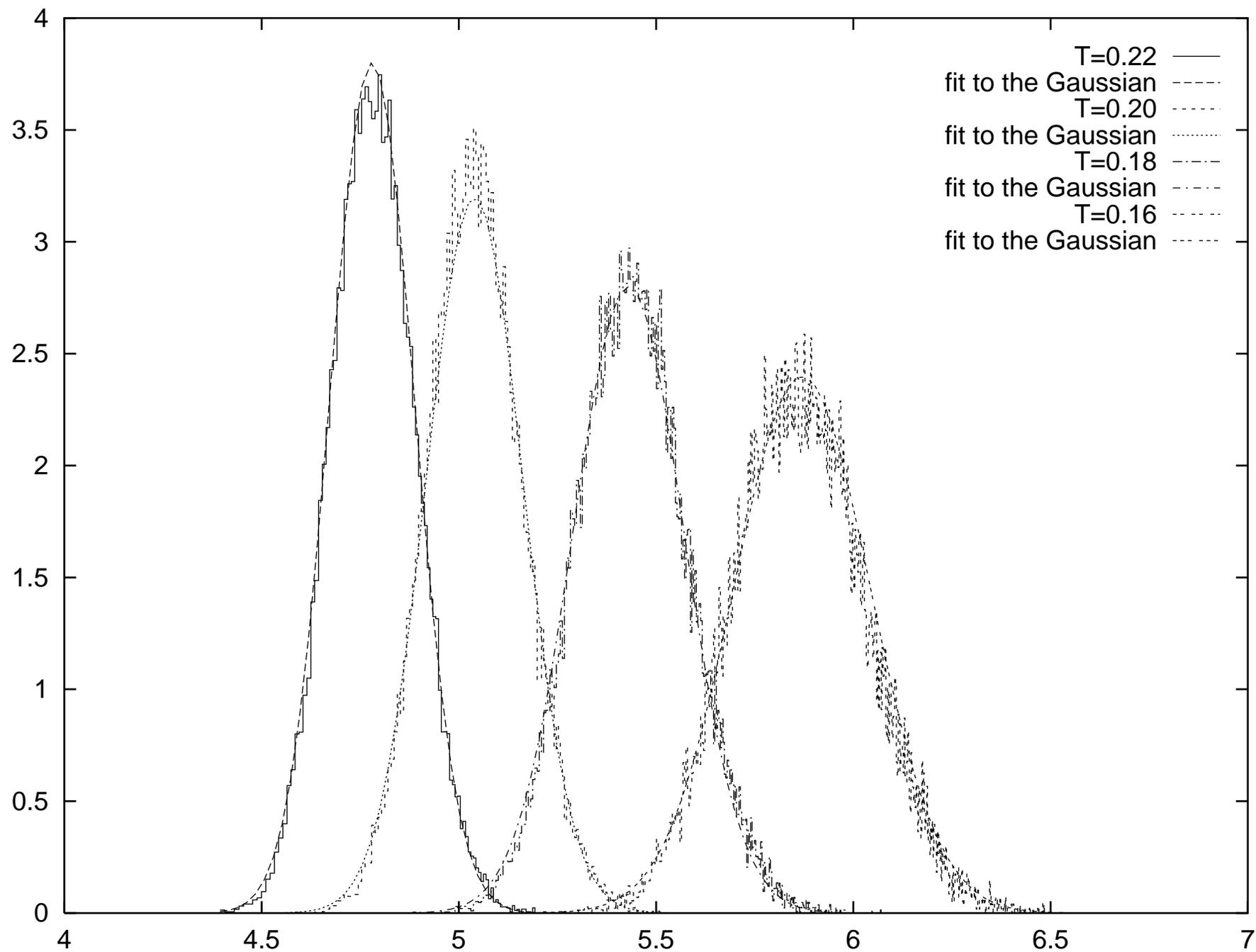
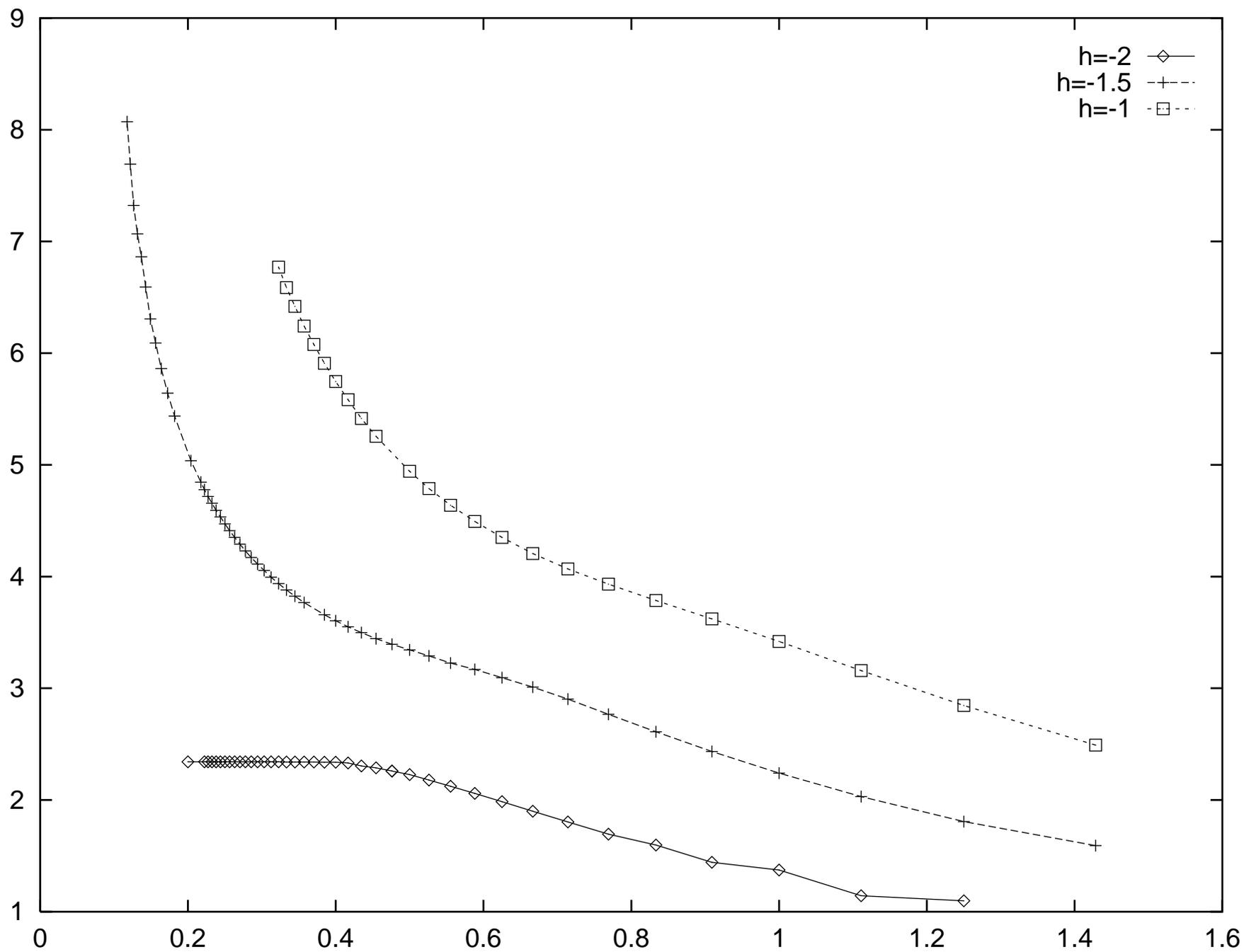
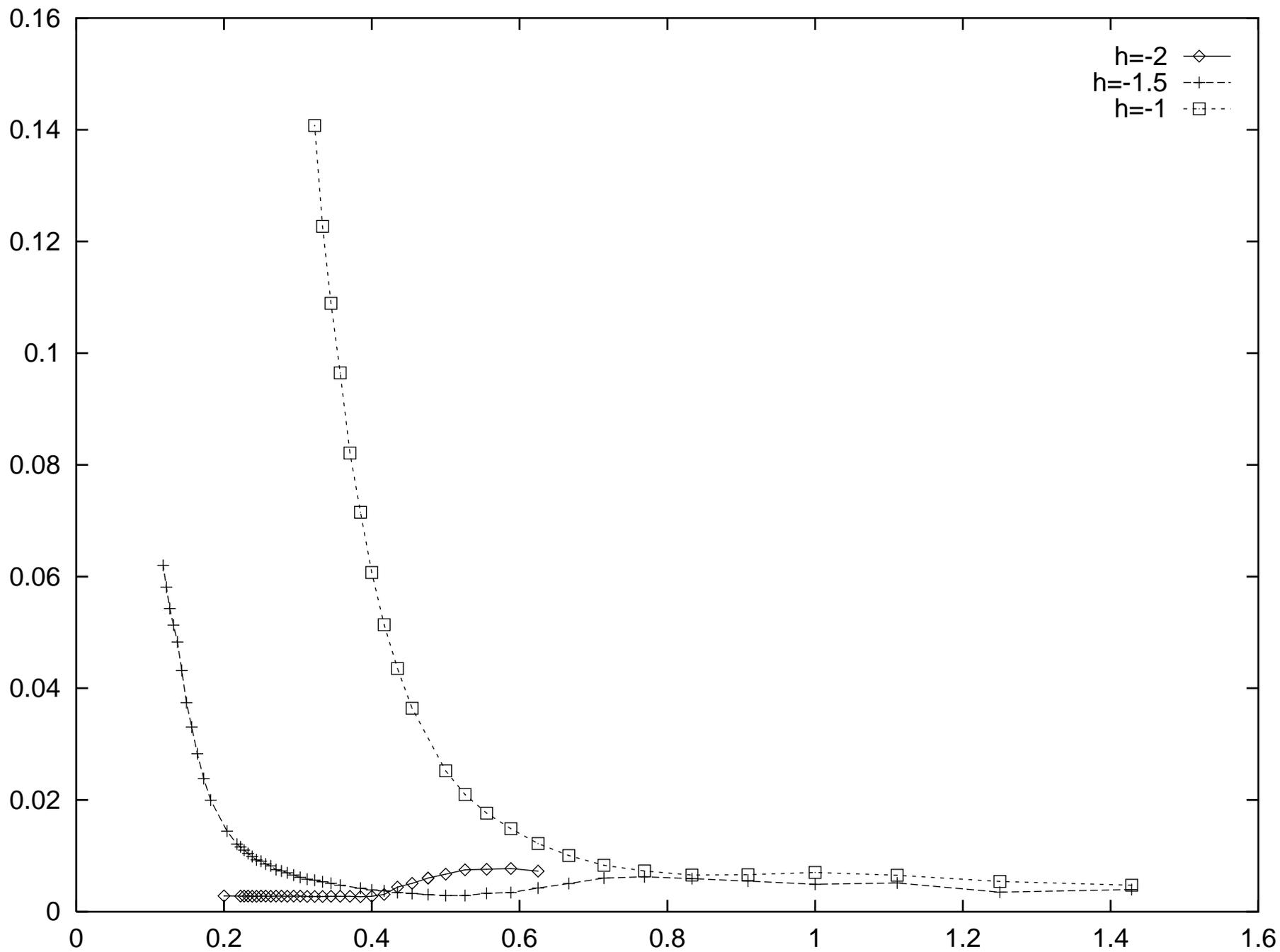
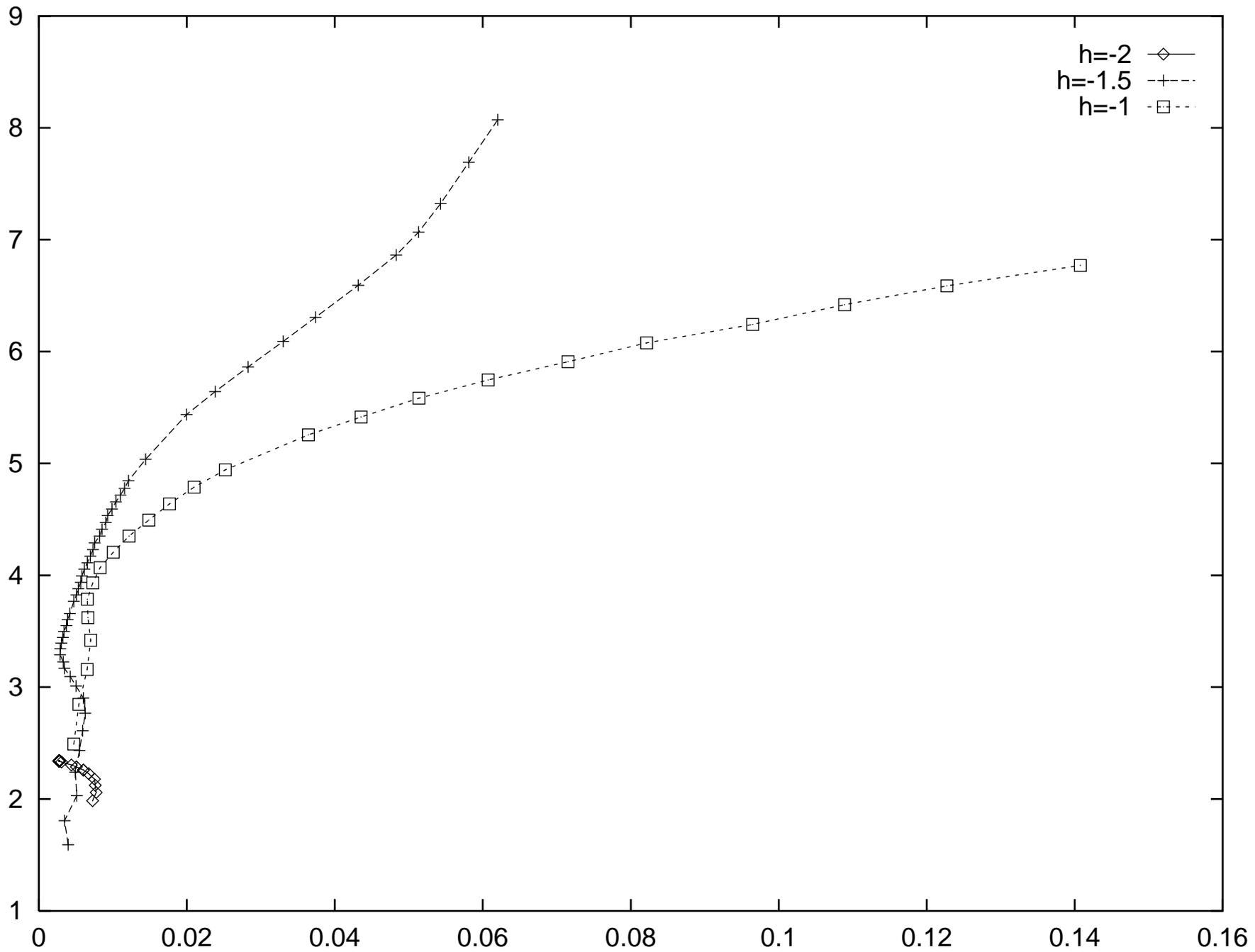
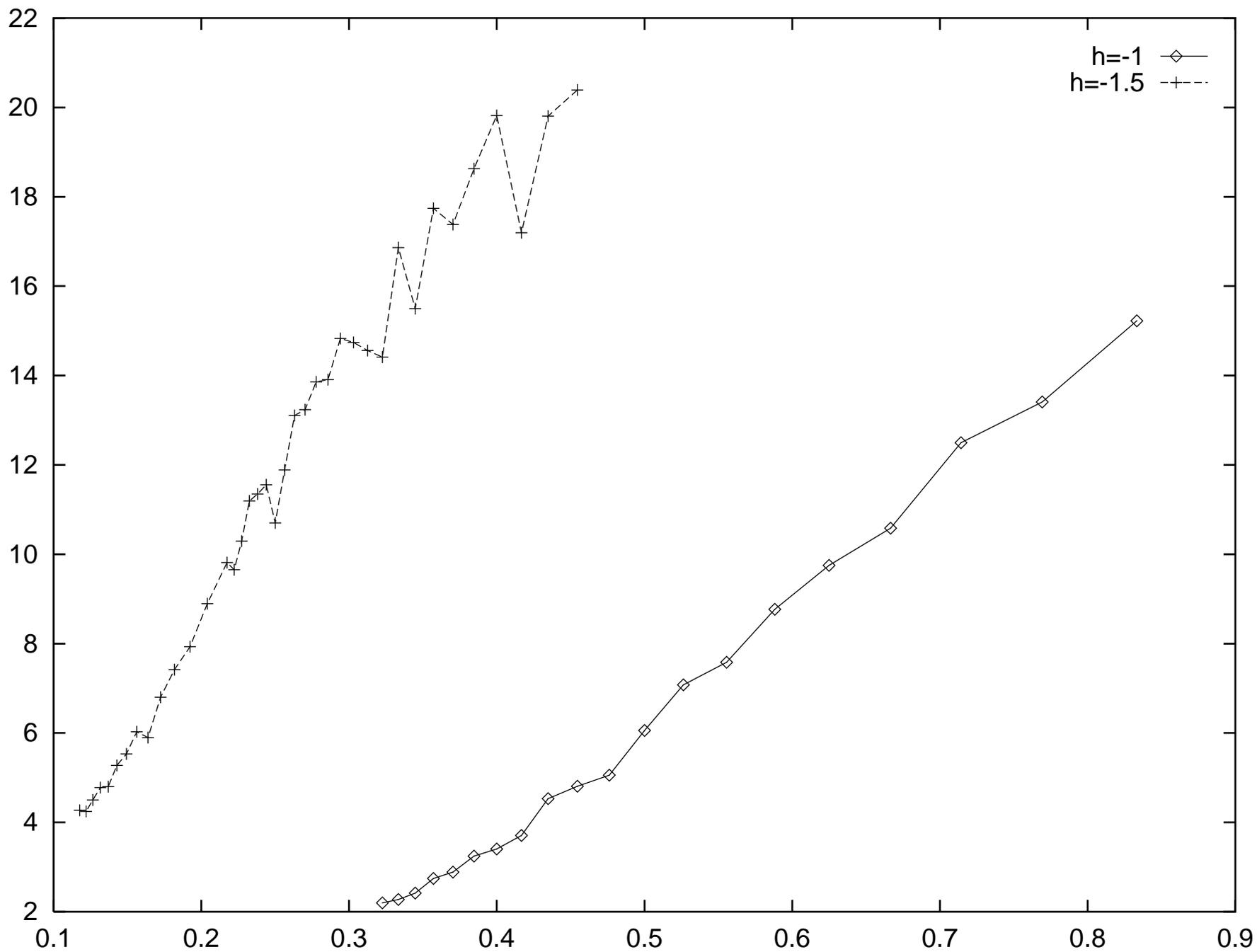
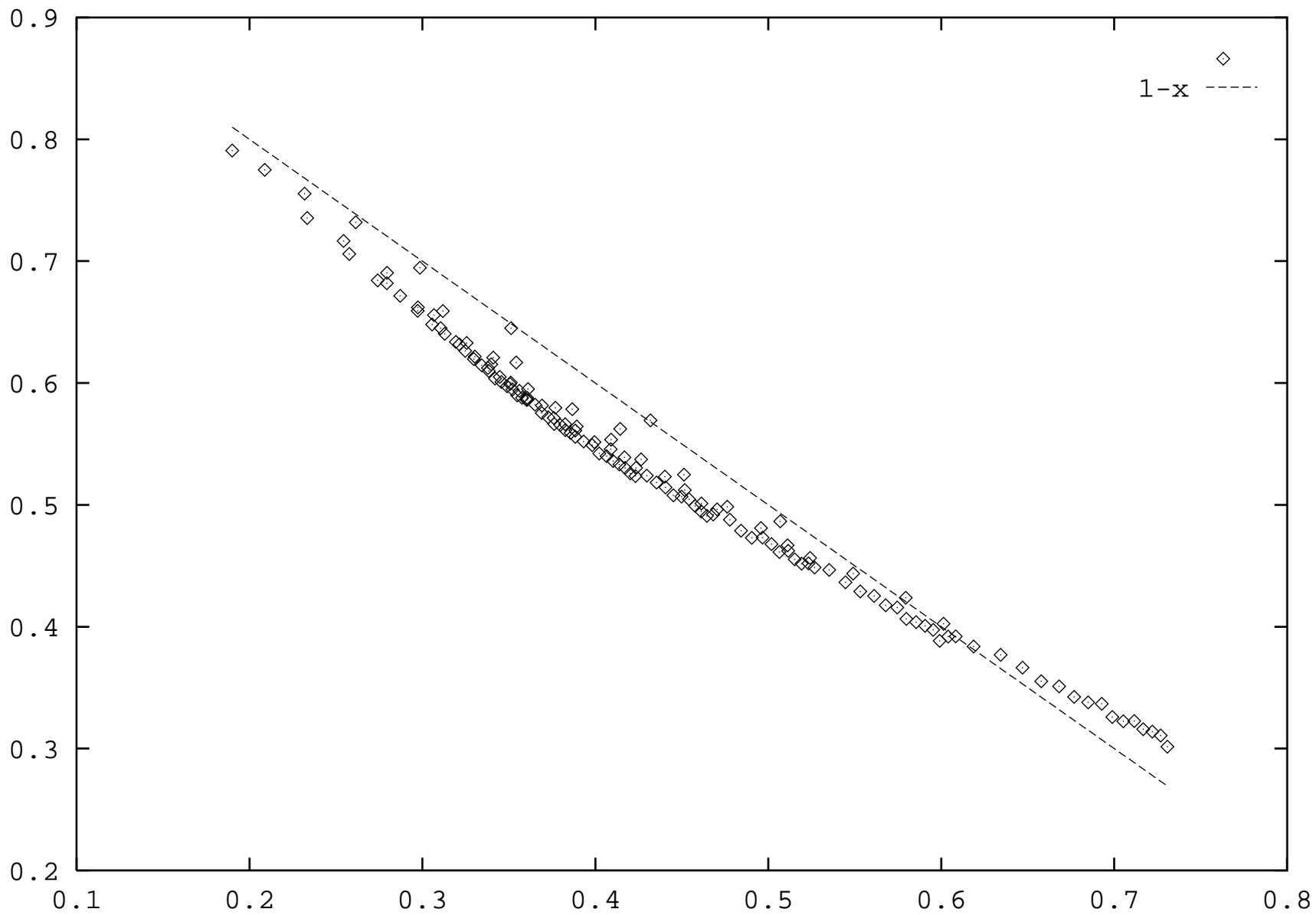
\end{document}